\documentclass{egs}
\usepackage{amsfonts}

\usepackage{amsmath}
\usepackage{graphicx}
\usepackage{epsfig}
\usepackage{subfigure}
\usepackage[scaled=.92]{helvet}
\usepackage{mathptmx}

 \printfigures
 \input{tcilatex}
 \DeclareMathAlphabet{\mathsf}{OT1}{cmss}{bx}{n}
 \DeclareMathAlphabet{\mathit}{OT1}{cmr}{bx}{it}
\begin{document}

\title{Passive tracer patchiness and particle trajectory stability in incompressible
two-dimensional flows}
\author{Francisco J. Beron-Vera, Mar\'{\i}a J. Olascoaga and Michael G. Brown}
\affil{RSMAS, University of Miami, Miami, Florida, USA}
\date{Manuscript version from \today.}
\journal{\NPG}
\firstauthor{Beron-Vera}
\proofs{F.~J. Beron-Vera\\RSMAS/AMP, UMiami\\4600 Rickenbacker Cswy.\\Miami, FL 33149 USA}
\offsets{F.~J. Beron-Vera\\RSMAS/AMP, UMiami\\4600 Rickenbacker Cswy.
\\Miami, FL 33149 USA}
\correspondence{F.~J. Beron-Vera (fberon@rsmas.\allowbreak
miami.edu)} \msnumber{1} \received{January 28, 2003}
\revised{\today} \accepted{} \runninghead{Beron-Vera et al.:
Tracer patchiness and particle stability} \firstpage{1}
\pubyear{2003} \pubvol{1} \pubnum{1}

\maketitle

\begin{abstract}
Particle motion is considered in incompressible two-dimensional
flows consisting of a steady background gyre on which an unsteady wave-like perturbation is
superimposed. A dynamical systems point of view that exploits the
action--angle formalism is adopted. It is argued and demonstrated
numerically that for a large class of problems one expects to observe a
mixed phase space, i.e., the occurrence of ``regular islands'' in an
otherwise ``chaotic sea.'' This leads to patchiness in the evolution of
passive tracer distributions. Also, it is argued and demonstrated
numerically that particle trajectory stability is largely controlled by the
background flow: trajectory instability, quantified by various measures of
the ``degree of chaos,'' increases on average with increasing
$\left|\mathrm{d}\omega/\mathrm{d}I\right|$, where $\omega (I)$ is
the angular frequency of the trajectory in the background flow and
$I$ is the action.
\end{abstract}%

\section{Introduction}

This paper deals with the kinematics of fluid particles in unsteady
incompressible flows on the Cartesian plane. Namely, we study properties of
trajectories $(x(t),y(t))$ that satisfy equations of the form%
\begin{mathletters}\label{sys}
\begin{equation}
\dot{x}=\partial _{y}\psi ,\quad \dot{y}=-\partial _{x}\psi ,
\end{equation}%
where the overdot stands for time derivative and $\psi (x,y,t)$ is the
streamfunction. Furthermore, we consider the latter to be split into a
steady background component and an unsteady perturbation component, i.e.,
\begin{equation}
\psi =\psi ^{(0)}(x,y)+\varepsilon \psi ^{(1)}(x,y,t),
\end{equation}%
\end{mathletters}%
where $\varepsilon $ is a dimensionless parameter. Equations (\ref{sys})
constitute a canonical Hamiltonian system with $\psi $ the Hamiltonian and $%
(x,y)$ the generalized coordinate--conjugate momentum pair.

Two related issues are addressed in this paper. First, we investigate a
cause of ``patchiness'' in passive tracer distributions, i.e., distributions
that are mostly vigorously stirred but include poorly stirred regions (Sect. %
\ref{pat}). Second, we study the influence of the background flow on
particle trajectory stability (Sect. \ref{sta}). Prior to discussing these
issues, the kinematic models that we use to illustrate our results are
briefly described in Sect. \ref{kin}. The conclusions of the paper are given
in Sect. \ref{con}.

\section{Kinematic models\label{kin}}

Two background flow structures in a region $[0,L]\times \lbrack 0,W]$ of the
$\beta $ plane are considered here. One is chosen to represent a large-scale
single-gyre wind-driven ocean circulation with streamfunction given by
\citep[][]{Stommel-66}%
\begin{equation*}
\text{S}:\psi ^{(0)}=a\left[ b\mathrm{e}^{b_{+}x}+(1-b)\mathrm{e}^{b_{-}x}-1%
\right] \sin \frac{\pi y}{W},
\end{equation*}%
where $a:=\tau W/(\pi \lambda D),$ $b:=(1-\mathrm{e}^{b_{-}L})/(\mathrm{e}%
^{b_{+}L}-\mathrm{e}^{b_{-}L})$, and $b_{\pm }:=-\frac{1}{2}\beta /\lambda
\pm \frac{1}{2}[(\beta /\lambda )^{2}+(\pi /W)^{2}]^{\frac{1}{2}}$. Here, $D$
is the depth, $\tau $ the wind stress amplitude (per unit density), and $%
\lambda $ the bottom friction. The other background streamfunction chosen
corresponds to solid body rotation,%
\begin{equation*}
\text{R}:\psi ^{(0)}=\frac{\omega _{\mathrm{R}}}{2}\left[ \left(
x-L/2\right) ^{2}+\left( y-W/2\right) ^{2}\right] .
\end{equation*}%
The reason for this highly idealized choice will be discussed below.
Parameter values used in our numerical work are listed in Table \ref{par}.

%
%
\begin{table}[tb]
\caption[]{Background flow parameters.}
\smallskip
\renewcommand{\arraystretch}{1}
\iftwocol{\small}{}
\label{par}

\begin{tabular}{ll}
\hline
Parameter & Value \\ \hline
\multicolumn{1}{c}{$L$} & $10$ \textrm{Mm} \\
\multicolumn{1}{c}{$W$} & $2\pi $ \textrm{Mm} \\
\multicolumn{1}{c}{$D$} & $200$ \textrm{m} \\
\multicolumn{1}{c}{$g$} & $9.8$ \textrm{m} \textrm{s}$^{-2}$ \\
\multicolumn{1}{c}{$f_{0}$} & $10^{-4}$ \textrm{s}$^{-1}$ \\
\multicolumn{1}{c}{$\beta $} & $10^{-11}$ \textrm{m}$^{-1}$\textrm{s}$^{-1}$
\\
\multicolumn{1}{c}{$\tau $} & $2\times 10^{-3}$ \textrm{m}$^{-2}$\textrm{s}$^{-2}$ \\
\multicolumn{1}{c}{$\lambda $} & $10^{-5}$ \textrm{s}$^{-1}$ \\
\multicolumn{1}{c}{$\omega _{\mathrm{R}}$} & $2\pi $ \textrm{y}$^{-1}$ \\
\hline
\end{tabular}

\end{table}%

\begin{figure*}[tb]
\centerline{
\subfigure[]{\includegraphics[width=5.5cm,clip=]{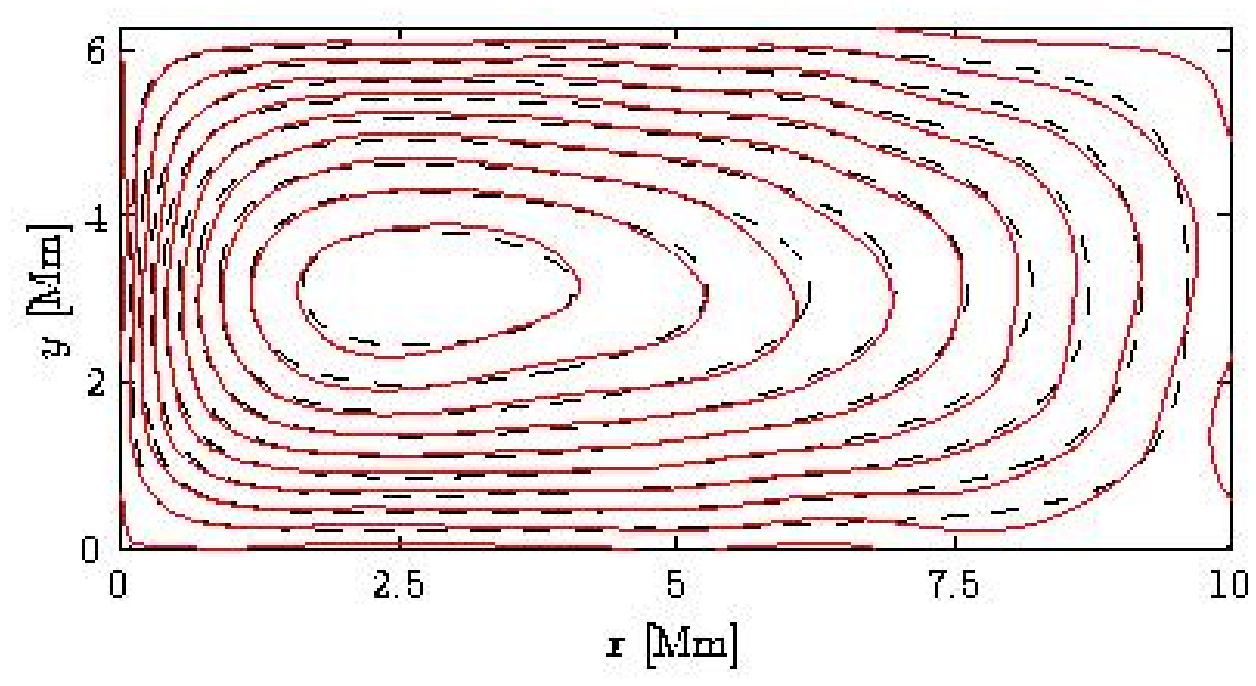}}\quad
\subfigure[]{\includegraphics[width=5.5cm,clip=]{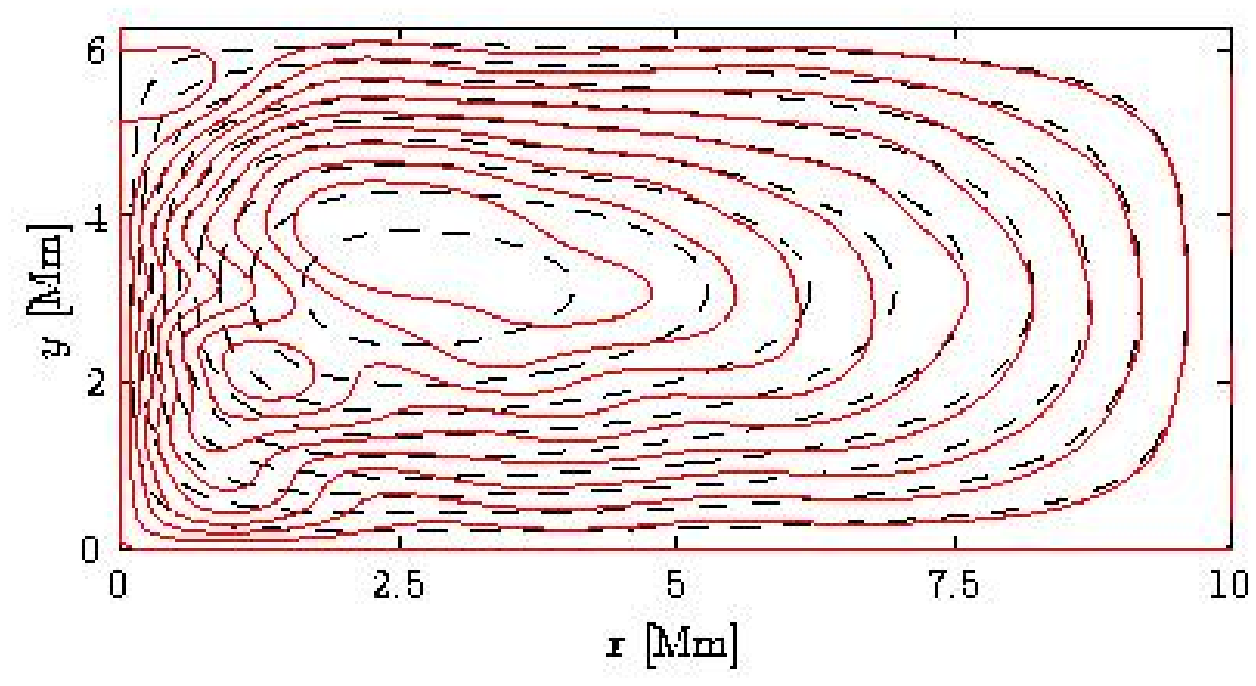}}\quad
\subfigure[]{\includegraphics[width=5.5cm,clip=]{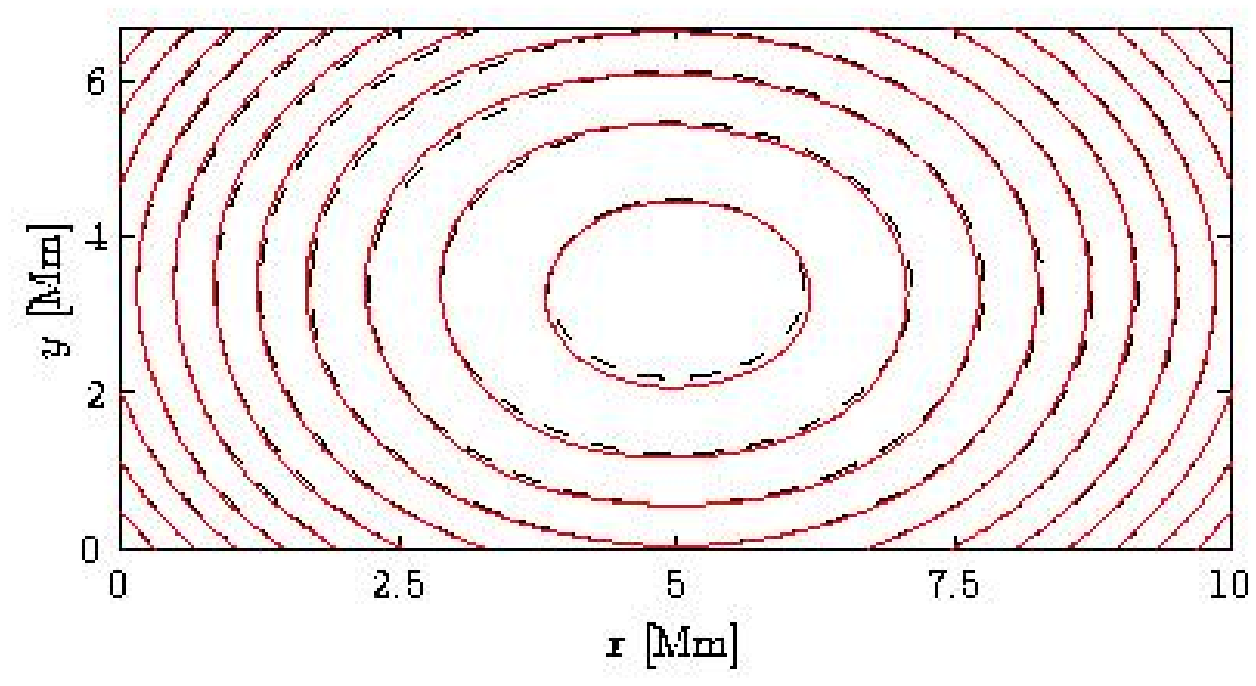}}}
\caption{Background flow streamlines (dashed lines) along with
streamlines corresponding to a snapshot of the total flow at
$t\approx 9$ $\mathrm{y}$ (solid lines). Panels a and b correspond
to background flow S with different wave-like perturbation fields
superimposed; panel c corresponds to background flow R with the
same perturbation that was used to produce panel a.\label{str}}
\end{figure*}%

The perturbation streamfunction is constructed by superposing standing
Rossby-like modes with a power-law spectrum, namely,
\begin{equation}
\psi ^{(1)}\hspace{-0.01in}=\hspace{-0.01in}a\sum_{k,l}A\mathrm{e}^{-\gamma
x}\hspace{-0.01in}\sin (kx\hspace{-0.01in}+\hspace{-0.01in}\phi _{k})\sin (ly%
\hspace{-0.01in}+\hspace{-0.01in}\phi _{l})\cos (\sigma t\hspace{-0.01in}+%
\hspace{-0.01in}\phi _{\sigma }),  \label{per}
\end{equation}%
where
\begin{gather*}
A(k,l):=\frac{\pi ^{2}(L^{-2}+W^{-2})}{k^{2}+l^{2}}, \\
\sigma (k,l):=-\frac{\beta k}{k^{2}+l^{2}+f_{0}^{2}/(gD)},
\end{gather*}%
and the $\phi (k,l)$'s are random numbers uniformly distributed between $0$
and $2\pi .$ Here, $Lk/\pi $ and $Wl/\pi \ $are positive integers; $\gamma $
is a constant; $f_{0}$ is the reference Coriolis parameter; and $g$ is the
acceleration of gravity.

Dashed lines in Figs. \ref{str}a,b and Fig. \ref{str}c are streamlines for
background flows S and R, respectively. Solid lines in these figures are
total flow streamlines corresponding to a snapshot of the flow at $t\approx
9 $ $\mathrm{y}$. The perturbation in each case involves $10\times 10=100$
modes. In Figs. \ref{str}a,c the perturbation has $\varepsilon =0.05$ and $%
\gamma =0.$ In Fig. \ref{str}b the amplitude of the $Lk/\pi =1=Wl/\pi $ mode
is set to zero, $\phi _{k}=0=\phi _{l}$ so the flow vanishes at the
boundary, $\varepsilon =0.25,$ and $\gamma =0.4$ $\mathrm{Mm}^{-1}$.

The flows used to produce Fig. \ref{str} and all of the numerical particle
trajectory simulations presented in this paper were chosen to illustrate
important aspects of Lagrangian dynamics; the flows are in many ways not
representative of realistic oceanic flows. We\emph{\ }note, however, that we
focus on flows with complicated time dependence, and that the strong
perturbations to the background are considered. In Fig. \ref{str}b, for
example, it is seen that the perturbation leads to the presence of an
eddy-like structure in the flow. Also, we note that in the flows that we
have described, particle trajectories are periodic in the limit of zero
perturbation strength with typical periods of about $1$ $\mathrm{y}$. Thus
in an integration time of $10$ $\mathrm{y}$ most trajectories will have made
approximately $10$ revolutions around the gyre. The phenomena described
below are not limited to gyre-scale flows. In general, the trends that we
describe should be evident after times in excess of a few periods of
particle revolution in any background gyre flow on which a perturbation
field with a broad band of frequencies is superimposed.

\section{Passive tracer patchiness\label{pat}}

In this section we present numerical evidence and a theoretical argument
that suggest that for a large class of systems of the form (\ref{sys}) phase
space $(x,y)$ should be partitioned into ``regular islands'' in a ``chaotic
sea.'' Such a mixed phase space leads to patchiness in passive tracer
distributions. Numerical results are presented for a time-periodic flow [$%
n=1 $ term in the sum in (\ref{per})] and subsequently for flows with
complicated time dependence ($n$ large).

\begin{figure}[tb]
\includegraphics[width=8cm,clip=]{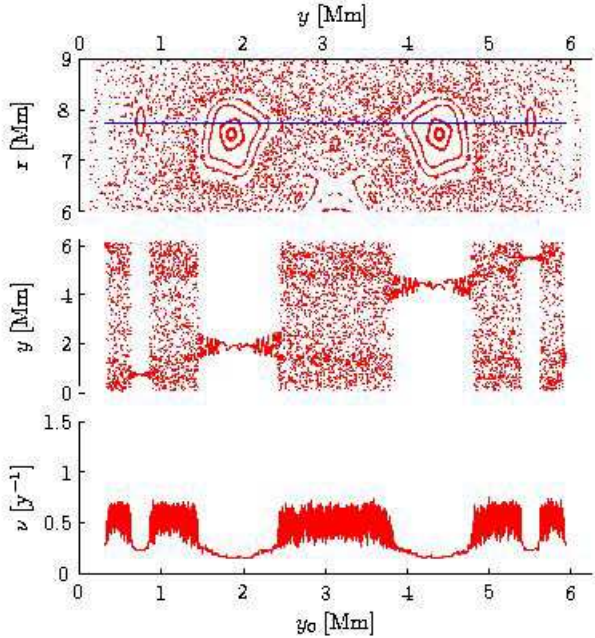}
\caption{Poincar\'e section (top), final vs. initial meridional
position (middle), and finite-time estimate of the Lyapunov
exponent as a function of initial meridional position (bottom).
The middle and lower plots were constructed by tracking $10^4$
particles ($\Delta y_0 \approx 5.5$ km) for a duration of 150 y in
background flow S with a time-periodic perturbation superimposed.
Particle initial positions fall on the horizontal line shown in
the top panel. The perturbation parameters chosen were
$\varepsilon =0.015$ , $\gamma =0,$ $kL/\pi =3=lW/\pi,$ $\phi $'s
$=0,$ and $2\pi /\sigma =0.25$ y.\label{poi}}
\end{figure}%

Figure \ref{poi} shows, for the time-periodic case ($n=1$), a Poincar\'{e}
section and, in the same environment, two additional trajectory diagnostics
whose applicability is not restricted to time-periodic flows. The Poincar%
\'{e} section was constructed by plotting the $(x,y)$ coordinates of several
trajectories at integer multiples of the period of the streamfunction; it
shows the usual mixture of ``regular islands'' in an otherwise ``chaotic
sea''
\citep[cf., e.g.,][]{Tabor-89}%
. The middle panel shows, for a dense set of trajectories with $x(0)=x_{0}$
fixed and $y(0)=y_{0}$ variable, a plot of $y$ vs. $y_{0}$ at a fixed value
of $t.$ The initial conditions chosen fall inside the region of the Poincar%
\'{e} section shown, and it is seen that both regular islands and the
chaotic sea evident in the Poincar\'{e} section can be identified in the $y$
vs. $y_{0}$ plot. The same structures can also be seen in the lower panel of
Fig. \ref{poi} which shows, for the same trajectories used to produce the
middle panel, finite time estimates of Lyapunov exponents (described in more
detail below), $\nu $ vs. $y_{0}.$ Plots of $y$ vs. $y_{0}$ and $\nu $ vs. $%
y_{0}$ are used below to distinguish between apparently regular and
apparently chaotic trajectories for flows with complicated ($n$ large) time
dependence i.e., in flows for which a Poincar\'{e} section cannot be
constructed.

\begin{figure*}[tb]
\centerline{
\subfigure[]{\includegraphics[width=5.5cm,clip=]{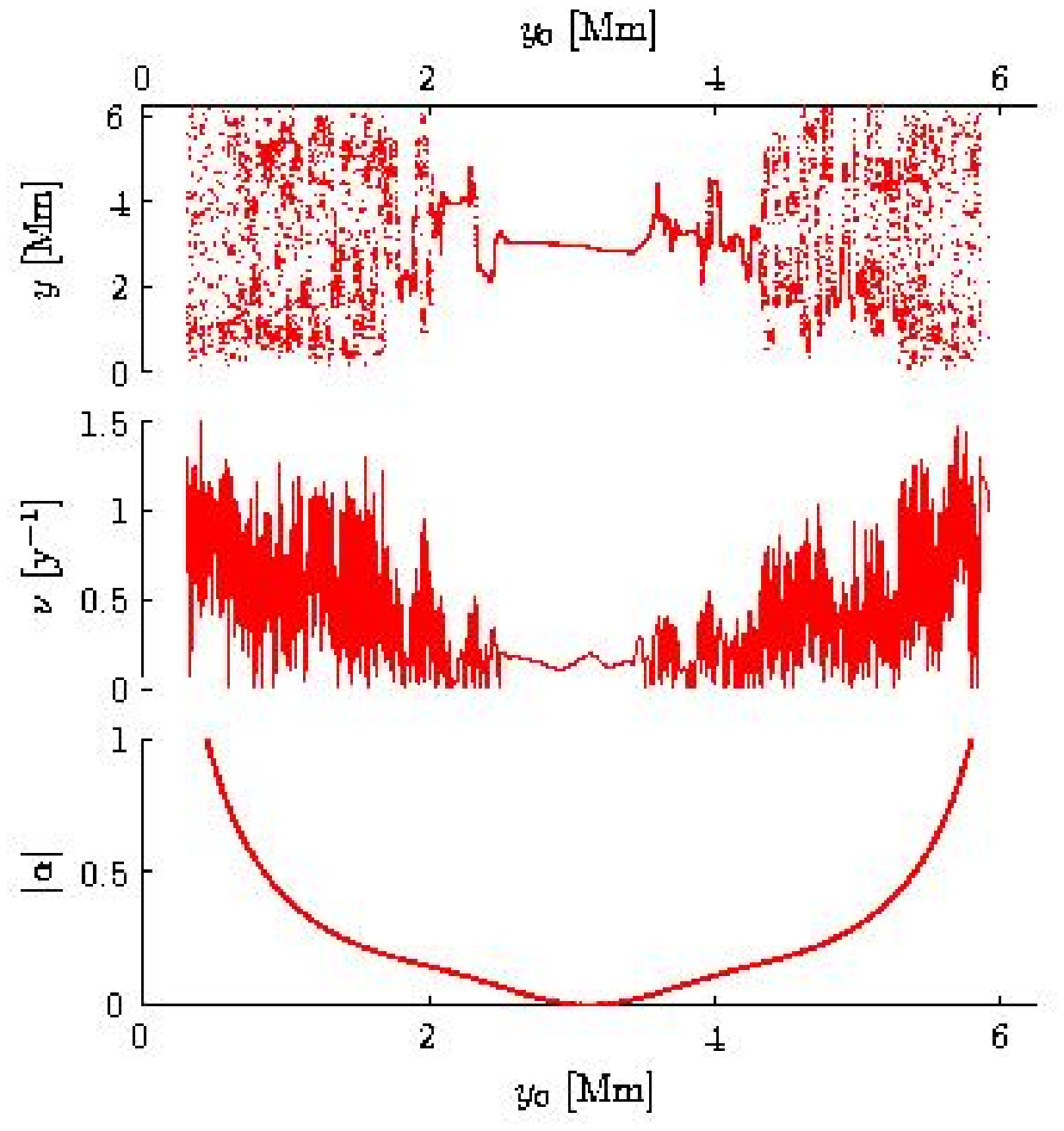}}\quad
\subfigure[]{\includegraphics[width=5.5cm,clip=]{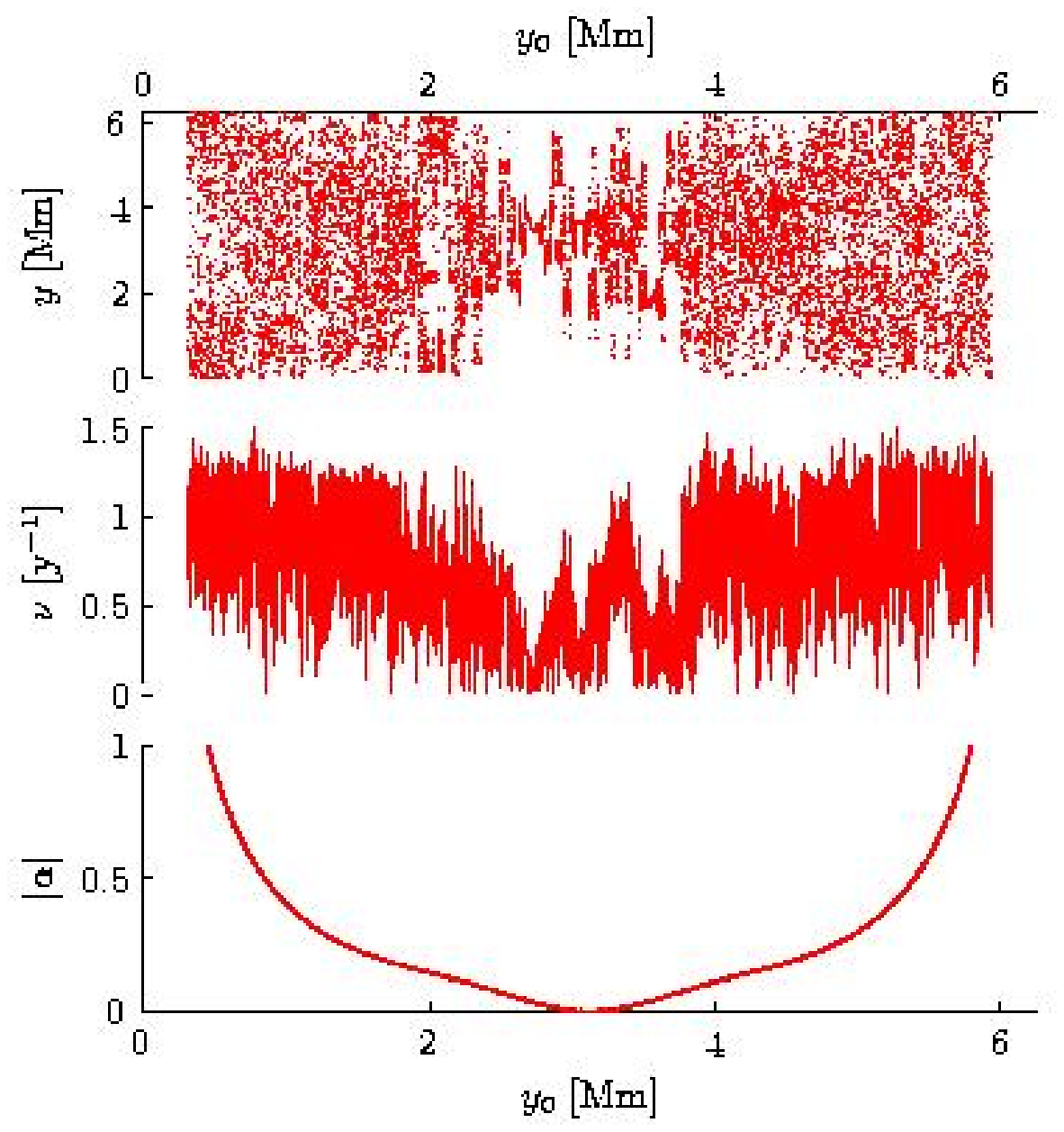}}\quad
\subfigure[]{\includegraphics[width=5.5cm,clip=]{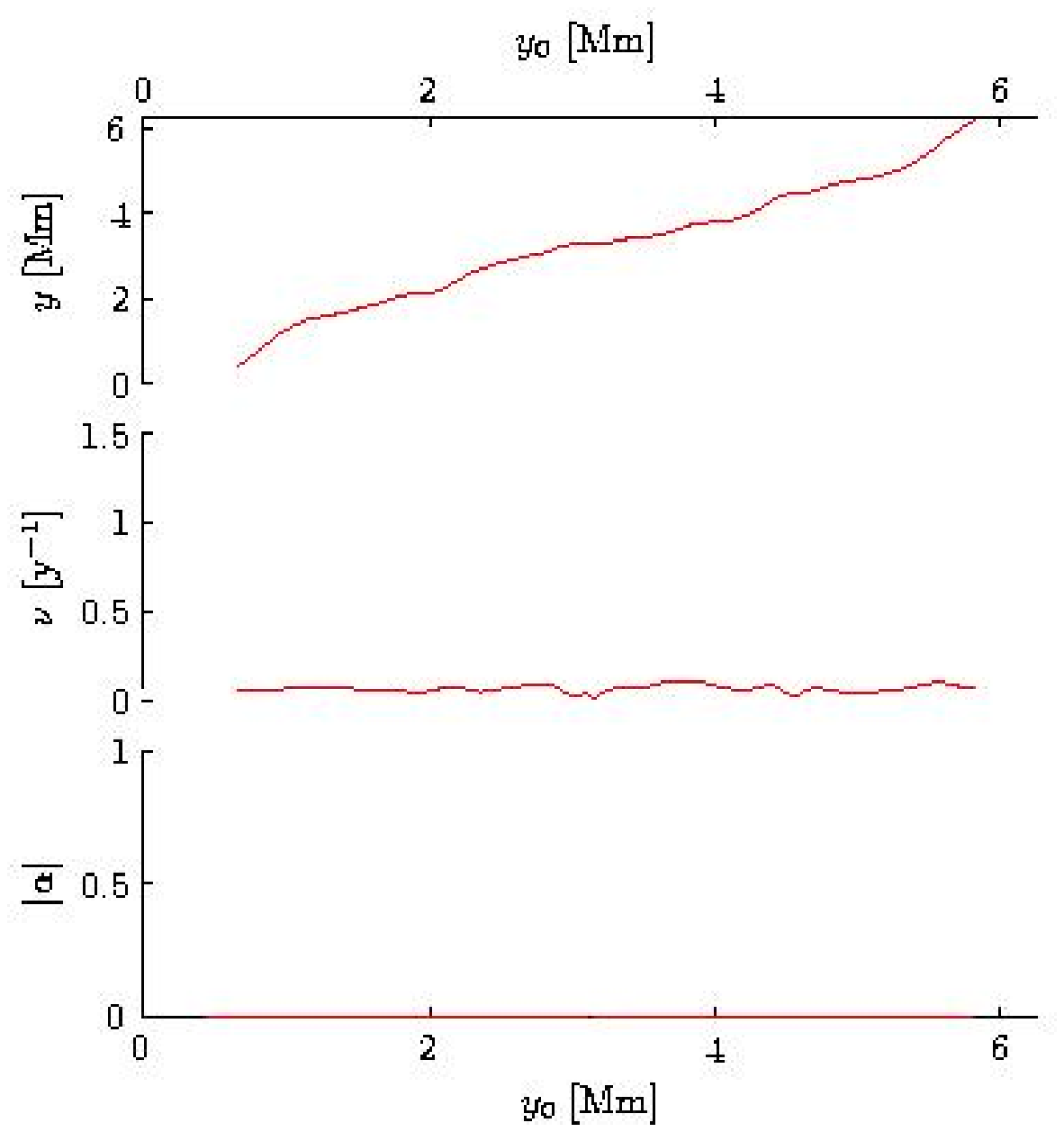}}}
\caption{Trajectory final meridional position $y$ (top panels),
finite-time estimate of the Lyapunov exponent $\nu$ (middle
panels), and absolute value of stability parameter $\alpha$
(bottom panel) as a function of initial meridional position.
Background and perturbation fields are as in Fig. \ref{str}. All
particles have initial longitudinal positions $x_0$ at the center
of the background gyre. The integration time is 12 y and $\Delta
y_0 \approx 5.5$ km.\label{alp}}
\end{figure*}

Figure \ref{alp} shows plots of $y$ vs. $y_{0}$ and $\nu $ vs. $y_{0}$ for
the nonperiodic flows used to produce Fig. \ref{str}. Trajectories in Fig. %
\ref{alp}b are generally more unstable than in Fig. \ref{alp}a. The enhanced
stability in Fig. \ref{alp}a is reflected in a relatively unstructured $%
y(y_{0})$ plot and smaller (on average) Lyapunov exponents than are seen in
Fig. \ref{alp}b. In both cases the background flow structure is the same;
the difference in the stability behavior is due to the difference in the
strength of the perturbation. As expected, trajectory instability is seen to
increase with increasing perturbation strength.

The difference in trajectory stability seen in Figs. \ref{alp}a,c has a
different explanation. The same perturbation was used in both cases, so this
cannot be the cause of the difference. The cause is the influence of the
background flow; this topic will be discussed in detail in the following
section.

\begin{figure*}[tb]
\centerline{
\subfigure[]{\includegraphics[width=7cm,clip=]{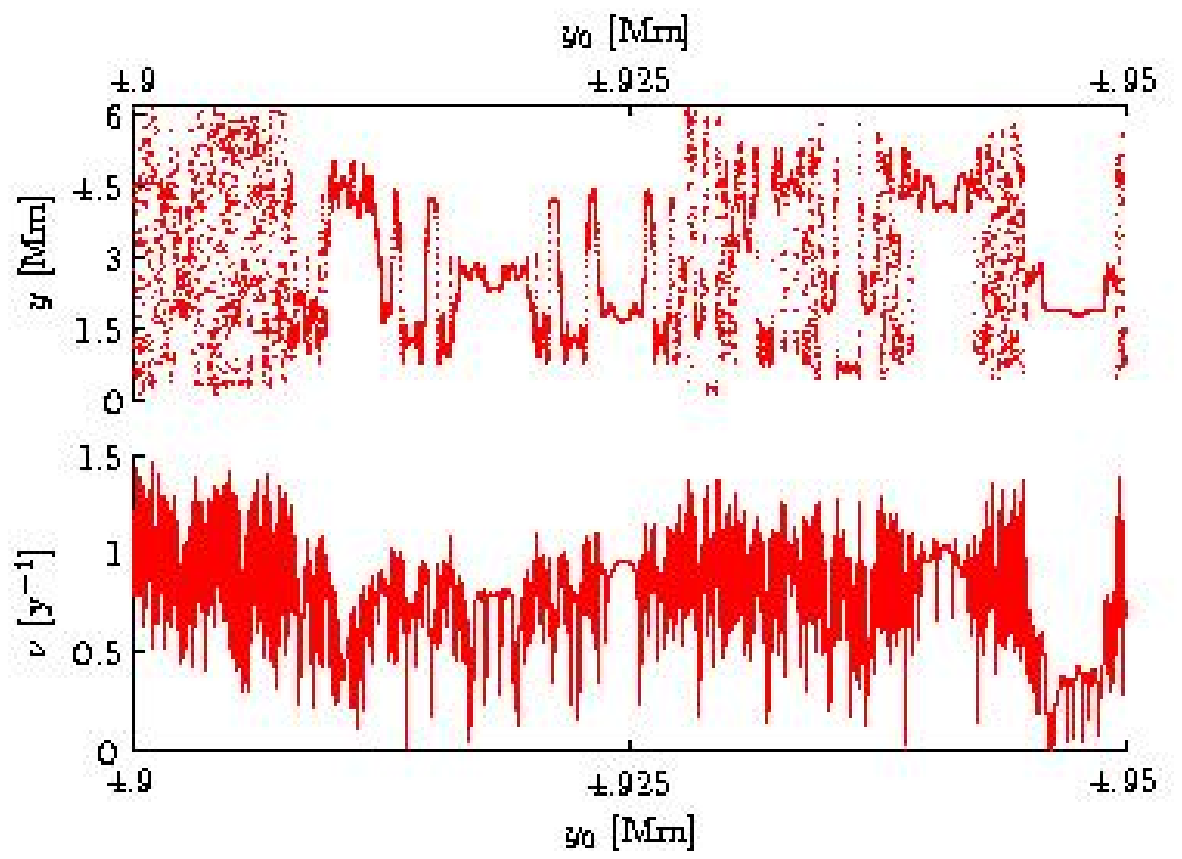}}\quad
\subfigure[]{\includegraphics[width=7cm,clip=]{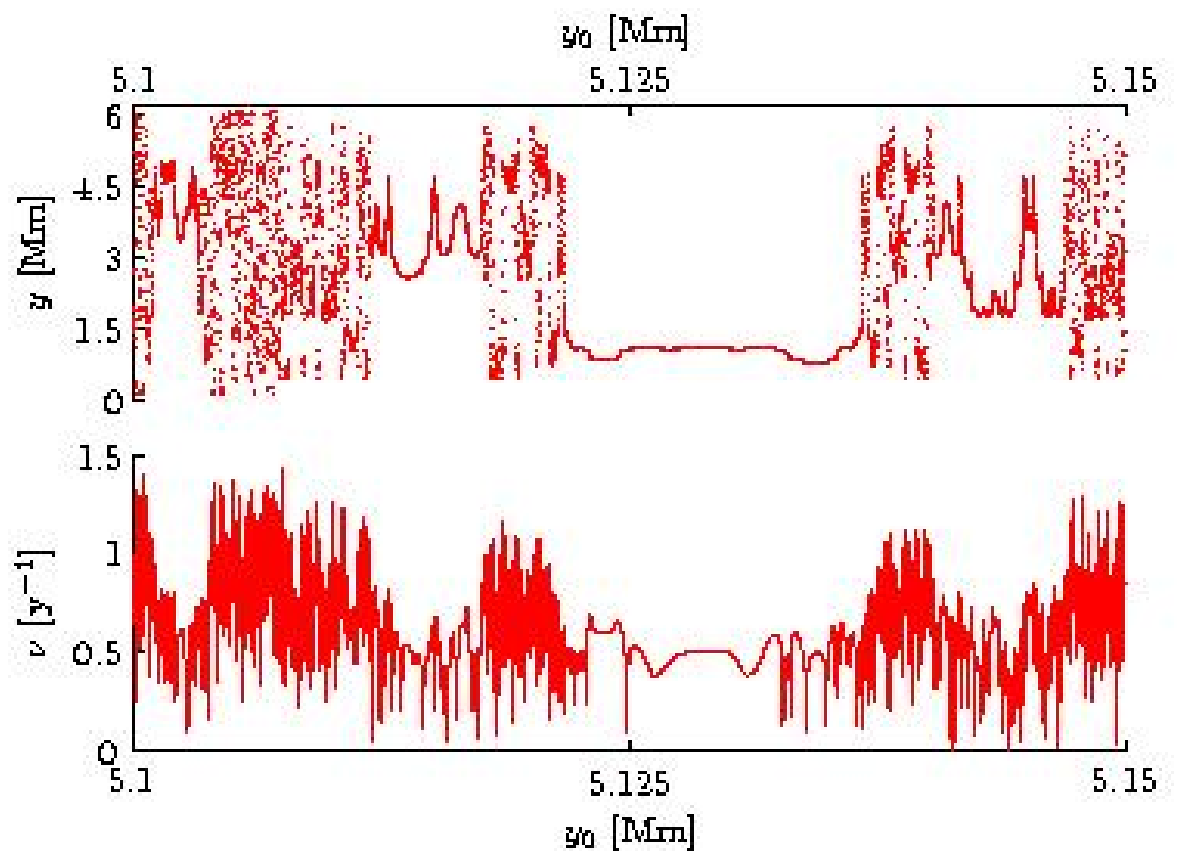}}}
\caption{Blow up of two portions of the top and middle panels of
Fig. \ref{alp}b using $\Delta y_0 \approx 5$ m.\label{alp-blo}}
\end{figure*}

We return our attention now to Fig. \ref{alp}b which corresponds to the
strongly perturbed flow shown in Fig. \ref{str}b. It is seen in Fig. \ref%
{alp}b that embedded among mostly chaotic trajectories are bands of
apparently nonchaotic trajectories. These nonchaotic bands are most readily%
\emph{\ }identified among the trajectories whose initial positions are near
the center of the gyre; the reason for this will be discussed in the
following section. Bands of nonchaotic trajectories far from the gyre center
are also present, however. This is seen in Fig. \ref{alp-blo} where two
regions of Fig. \ref{alp}b are blown up. These apparently nonchaotic bands
of trajectories are the counterparts of the ``regular islands'' seen in Fig. %
\ref{poi}. Trajectories in these bands diverge only very slowly (power law
dependence on time) from neighboring trajectories while chaotic trajectories
diverge at an exponential rate from neighboring trajectories. The nonchaotic
regions of flows are important in applications because they correspond to
regions where the concentration of a passive tracer will remain high for a
long duration. The existence of these regions leads to a large variance of
tracer concentration or ``tracer patchiness''
\citep[cf., e.g.,][]{Pasmanter-88,Malhotra-Mezic-Wiggins-98}%
.

\begin{figure*}[tb]
\centerline{
\subfigure[]{\includegraphics[width=5.5cm,clip=]{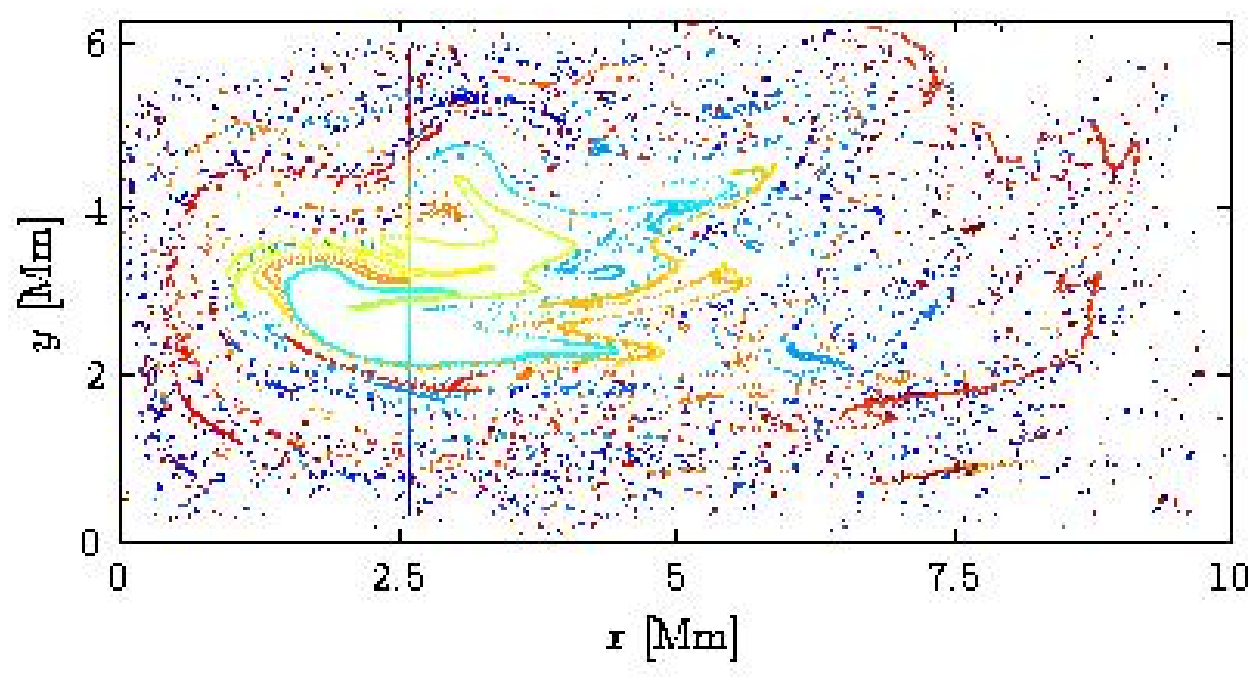}}\quad
\subfigure[]{\includegraphics[width=5.5cm,clip=]{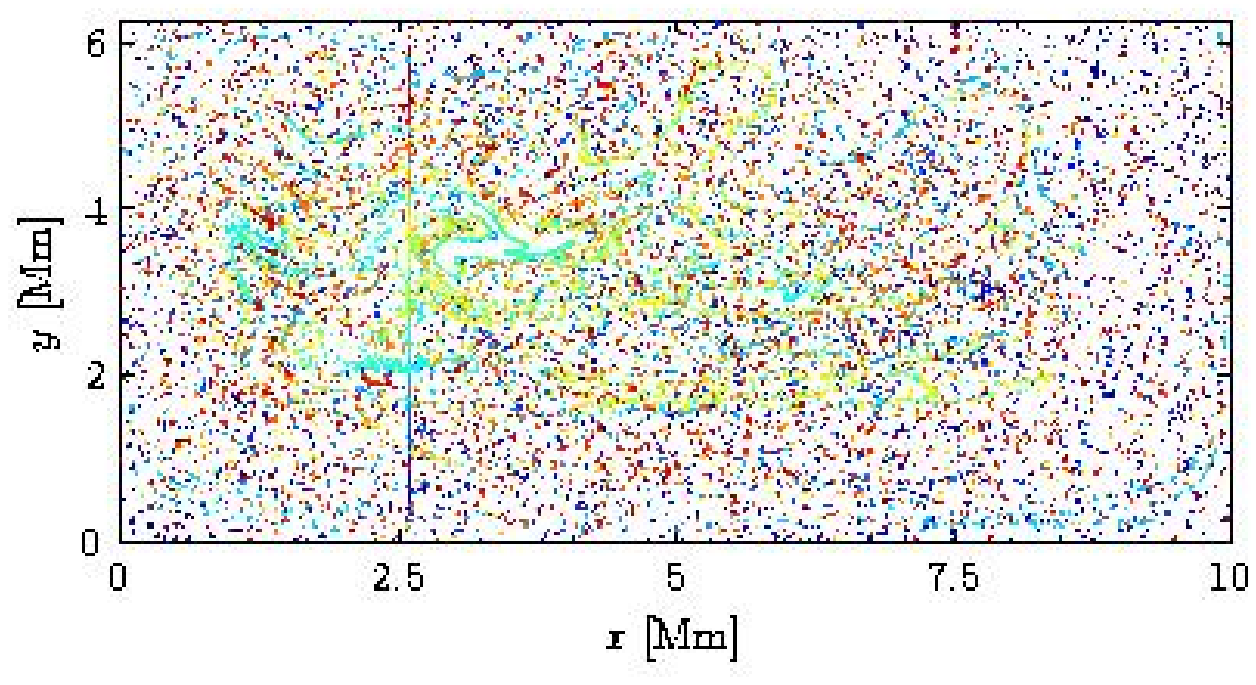}}\quad
\subfigure[]{\includegraphics[width=5.5cm,clip=]{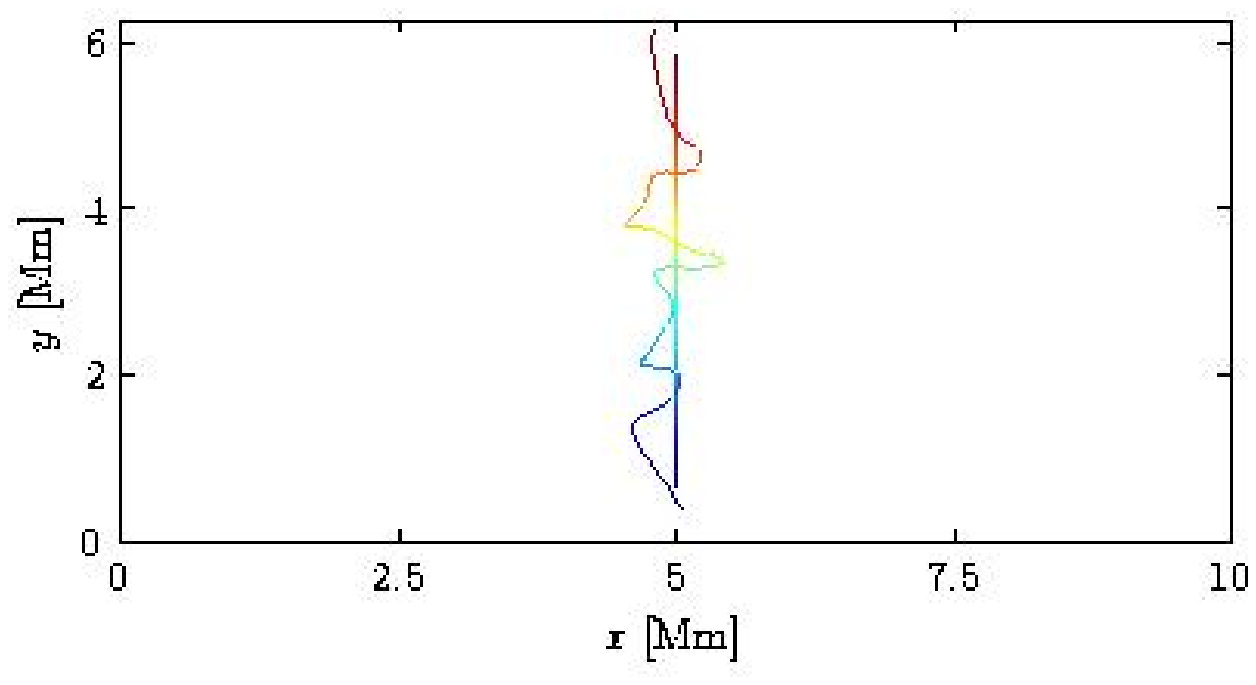}}}
\caption{Initial position (straight segment) and position after 12
y of a discretely sampled ($\Delta y_0 \approx 5.5$ km) material
line of fluid advected by the total (background plus perturbation)
flows of Fig. \ref{str}. Colors indicate the initial meridional
position of particles.\label{lag}}
\end{figure*}

\begin{figure*}[tb]
\centerline{
\subfigure[]{\includegraphics[width=7cm,clip=]{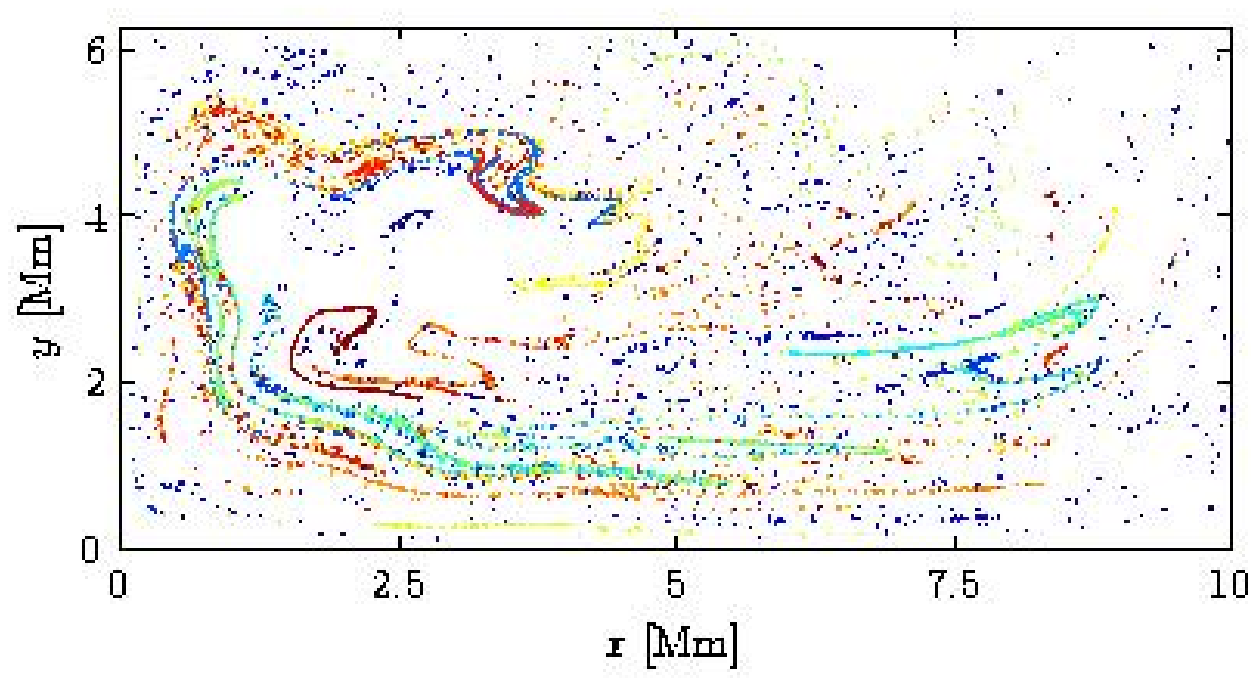}}\quad
\subfigure[]{\includegraphics[width=7cm,clip=]{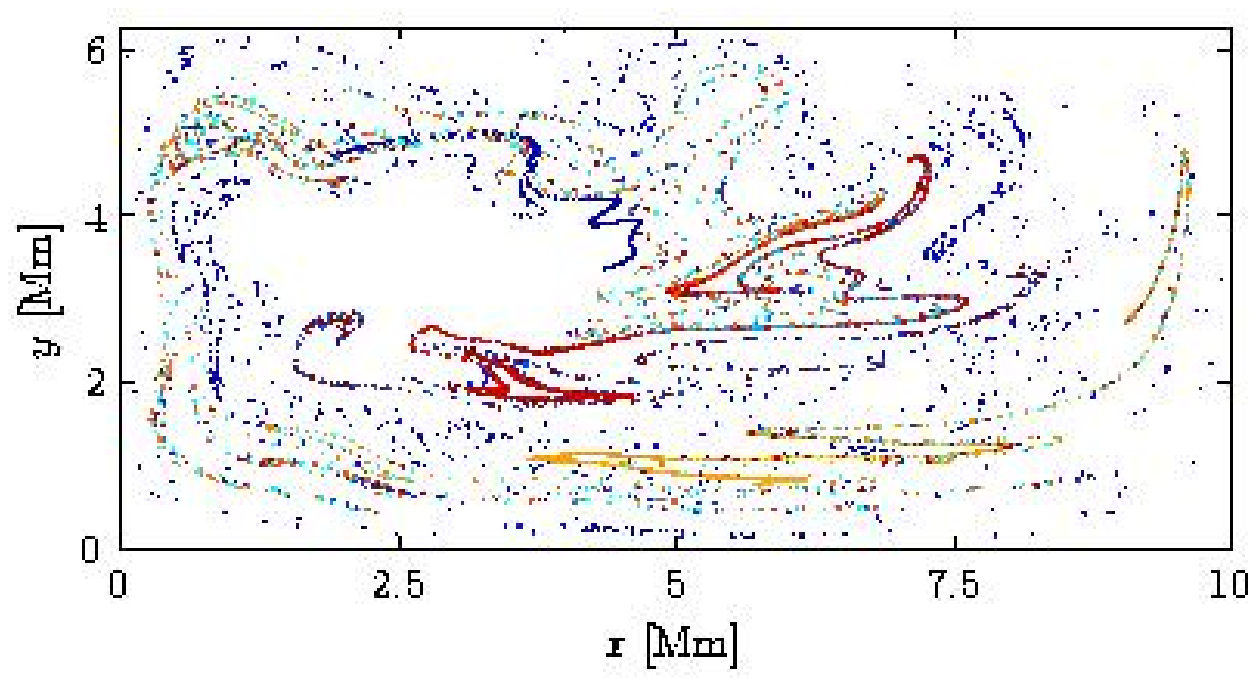}}}
\caption{Blow up of two portions of Fig. \ref{lag}b corresponding
to trajectories shown in Figs. \ref{alp-blo}a,b; here $\Delta y_0
\approx 5$ m. Colors indicate the initial meridional position of
particles; the initial material lines are so short that they
appear as dots.\label{lag-blo}}
\end{figure*}

Another way to visualize passive tracer patchiness is offered in Figs. \ref%
{lag} and \ref{lag-blo}. Both figures show discrete samples of a material
line of fluid at $t=0$ (vertical line segments in the figures) and at $t=12$
$\mathrm{y,}$ in the environments shown in Fig. \ref{str}. Initial
conditions in Fig. \ref{lag} are as in Fig. \ref{alp}, whereas those in Fig. %
\ref{lag-blo} are as in Fig. \ref{alp-blo}. These figures again show that
while most of the initial material line segment is vigorously stirred, there
are small portions of the initial segment, corresponding to the island-like
structures seen in Figs. \ref{alp} and \ref{alp-blo}, that are poorly
stirred.

We turn our attention now to explaining the occurrence of island-like
structures in Figs. \ref{poi}--\ref{alp-blo}. First, we note that in the
background flow, particle motion is describable using action--angle
variables, reviewed below, and trajectories fall on tori. For perturbed
systems with periodic time dependence, as in Fig. \ref{poi}, it is
well-known that particle trajectory dynamics are constrained by the KAM
theorem
\citep[cf., e.g.,][]{Arnold-89}
which guarantees that for sufficiently small $\varepsilon $ some of the
original tori---and associated nonchaotic motion---are preserved. Related
theoretical results, generally known as KAM theory
\citep[cf., e.g.,][]{Tabor-89}%
, describe how the nonsurviving tori break up to form chains of ``islands''
surrounded by a ``chaotic sea'' as seen in Fig. \ref{poi}. For a large
perturbation strength $\varepsilon $ all of the original tori will have been
broken up, but the secondary islands that are formed in the process are
robust and persist even when the magnitude of the perturbation exceeds that
of the background flow. It has been shown
\citep[][]{Brown-98,Beigie-etal-91}
that for multiply-periodic perturbations the situation is essentially the
same as for perturbations with simple periodic time dependence. This follows
from the observation that (\ref{sys}), with $\psi (x,y,$\emph{\ }$\sigma
_{1}t,\cdots ,\sigma _{n}t)$ where $\sigma _{i}t$ is defined modulo $2\pi $,
can be transformed to an autonomous Hamiltonian system with a bounded phase
space with\emph{\ }$(n+1)$ degrees of freedom that is constrained by $n$
integrals. KAM theory (the KAM\ theorem and related results) applies to the
transformed system, so phase space is generically partitioned into
nonintersecting regular and chaotic regions. A Poincar\'{e} section could,
in principle, be constructed for such a system by using a multiple slicing
technique
\citep[cf.][]{Parker-Chua-89}
but slicing is practical only when $n=1$. The significance of the extension
of KAM\ theory to multiply-periodic systems is that in the system defined by
(\ref{sys}) with $\psi ^{(1)}$ given by (\ref{per}), phase space $(x,y)$ is
expected to be partitioned into ``regular islands'' in an otherwise
``chaotic sea.'' The numerical evidence presented in Figs. \ref{alp}--\ref%
{lag-blo} supports this expectation.

The coexistence of regular and chaotic fluid particle trajectories in
mesoscale and large-scale oceanic flows has been suggested in some analyses
of surface drifters and submerged floats
\citep[][]{Osborne-etal-86a,Osborne-etal-89,Richardson-etal-89,Brown-Smith-90}%
. The preceding discussion provides an explanation of the underlying physics.

\section{Particle trajectory stability\label{sta}}

In this section we describe the important influence of the background flow
on particle trajectory stability that was mentioned above in our discussion
of Figs. \ref{alp}a,c. The ideas presented here apply to any canonical
Hamiltonian system in which the Hamiltonian consists of a superposition of
an integrable component and a nonintegrable perturbation. Other applications
are described in Beron-Vera and Brown (2003a,b). \nocite%
{Beron-Brown-03,Beron-Brown-03b}

The explanation of this behavior makes use of the action--angle description
of the motion of particles in the background flow
\citep[cf., e.g.,][]{Abdullaev-Zaslavsky-91}%
. Let
\begin{equation*}
I:=\frac{1}{2\pi }\oint \mathrm{d}x\,Y(x;\psi ^{(0)}),
\end{equation*}%
where $Y$ is the meridional coordinate of an isoline of $\psi ^{(0)}$, be
the action variable, and consider the canonical transformation $(x,y)\mapsto
(\vartheta ,I),$ defined implicitly by
\begin{equation*}
y=\partial _{x}G,\;\vartheta =\partial _{I}G,\;G(x,I):=\int \mathrm{d}%
x\,Y(x;\psi ^{(0)}),
\end{equation*}%
where $\vartheta $ is the angle variable. According to the above
transformation,%
\begin{equation*}
\psi (x,y,t)\mapsto \bar{\psi}^{(0)}(I)+\varepsilon \bar{\psi}%
^{(1)}(I,\vartheta ,t)
\end{equation*}%
and Eqs. (\ref{sys}a) take the form
\begin{equation}
\dot{I}=-\varepsilon \partial _{\vartheta }\bar{\psi}^{(1)},\quad \dot{%
\vartheta}=\omega +\varepsilon \partial _{I}\bar{\psi}^{(1)},  \label{act}
\end{equation}%
where
\begin{equation}
\omega (I):=\mathrm{d}\bar{\psi}^{(0)}/\mathrm{d}I.  \label{ome}
\end{equation}

When $\varepsilon =0$, Eqs. (\ref{act}), which have one degree of freedom,
are autonomous and the corresponding Hamiltonian, $\bar{\psi}^{(0)},$ is an
integral of motion that constrains the dynamics. As a consequence, the
equations can be solved by quadratures and the motion is periodic with
angular frequency $\omega $. Namely, $I=I_{0}$ and $\vartheta =\vartheta
_{0}+\omega t$ $\func{mod}2\pi $, where $I_{0}$ and $\vartheta _{0}$ are
constants. Every solution curve is thus a line that winds around an
invariant one-dimensional torus $\{I_{0}\}\times T^{1}\subset \mathbb{R}%
\times T^{1}$, whose representation in $(x,y)$-space is the closed curve
given by the isoline $\psi ^{(0)}=\bar{\psi}^{(0)}(I_{0}).$

With the perturbation term, the corresponding Hamiltonian, $\bar{\psi}%
^{(0)}+\varepsilon \bar{\psi}^{(1)},$ is no longer an integral of motion
(the equations are nonautonomous) and the system may be\ sensitive to initial%
\textit{\ }conditions, thereby leading to\ chaotic motion. The distinction
between regular and chaotic trajectories is commonly quantified by the%
\textit{\ }Lyapunov exponent
\citep[cf., e.g.,][]{Parker-Chua-89}%
, a measure of the rate at which neighboring trajectories diverge,\
\begin{equation}
\nu _{\infty }:=\lim\limits_{t\rightarrow \infty }\dfrac{1}{t}\ln |\nu ^{%
\mathsf{Q}}|\text{\textit{,}}  \label{lya}
\end{equation}%
where $\nu ^{\mathsf{Q}}(t)$ is the largest of the two eigenvalues of the
so-called stability matrix $\mathsf{Q}(t)$, which is given by%
\renewcommand{\arraystretch}{.9}
\renewcommand{\arraycolsep}{.02in}%
\begin{equation*}
\mathsf{Q}:=\left[
\begin{array}{cc}
\partial _{I_{0}}I & \partial _{\vartheta _{0}}I \\
\partial _{I_{0}}\vartheta & \partial _{\vartheta _{0}}\vartheta%
\end{array}%
\right] .
\end{equation*}%
Because of the area preservation property of Eqs. (\ref{sys}a) or (\ref{act}%
) the product of the two eigenvalues of $\mathsf{Q}$ is unity, so there is
no loss of generality in considering only the largest eigenvalue. Each
column of $\mathsf{Q}$ corresponds to a vector perturbation $(\delta
I,\delta \vartheta )$ to a trajectory in the nonautonomous system (\ref{act}%
), and satisfies the so-called variational equations,%
\begin{equation}
\left(
\begin{array}{c}
\delta \dot{I} \\
\delta \dot{\vartheta}%
\end{array}%
\right) \hspace{-0.02in}\hspace{-0.02in}=\hspace{-0.02in}\hspace{-0.02in}%
\left[
\begin{array}{cc}
0 & 0 \\
\omega ^{\prime } & 0%
\end{array}%
\right] \hspace{-0.02in}\hspace{-0.02in}\left(
\begin{array}{c}
\delta I \\
\delta \vartheta%
\end{array}%
\right) \hspace{-0.02in}\hspace{-0.02in}+\hspace{-0.02in}\varepsilon \hspace{%
-0.02in}\hspace{-0.02in}\left[
\begin{array}{rr}
-\partial _{I\vartheta }\bar{\psi}^{(1)} & -\partial _{\vartheta \vartheta }%
\bar{\psi}^{(1)} \\
\partial _{II}\bar{\psi}^{(1)} & \partial _{I\vartheta }\bar{\psi}^{(1)}%
\end{array}%
\right] \hspace{-0.02in}\hspace{-0.02in}\left(
\begin{array}{c}
\delta I \\
\delta \vartheta%
\end{array}%
\right) ,  \label{var}
\end{equation}%
where $\omega ^{\prime }:=\mathrm{d}\omega /\mathrm{d}I.$ Equations (\ref%
{var}) and (\ref{act}) constitute a system of four coupled equations.%
\renewcommand{\arraystretch}{1}
\renewcommand{\arraycolsep}{.03in}%

Variational equations that describe the growth of perturbations using
Cartesian coordinates $(\delta x,\delta y)$ have the same form as (\ref{var}%
) except that in the Cartesian form all four elements of the first matrix on
the r.h.s. of (\ref{var}) are generally nonzero. Our numerical finite-time
Lyapunov exponent estimates are based on the Cartesian equivalent of Eqs. (%
\ref{var}) and (\ref{act}), which is generally more convenient for numerical
calculations. We have chosen to show the $(\delta I,\delta \vartheta )$ form
of these equations to highlight the important role played by $\omega
^{\prime }$. An example of a closely related study which does not exploit
action--angle variables, and which consequently overlooks the critical
importance of $\omega ^{\prime }$, is
\citet{Richards-etal-95}%
.

\begin{figure*}[tb]
\centerline{
\subfigure[]{\includegraphics[width=7cm,clip=]{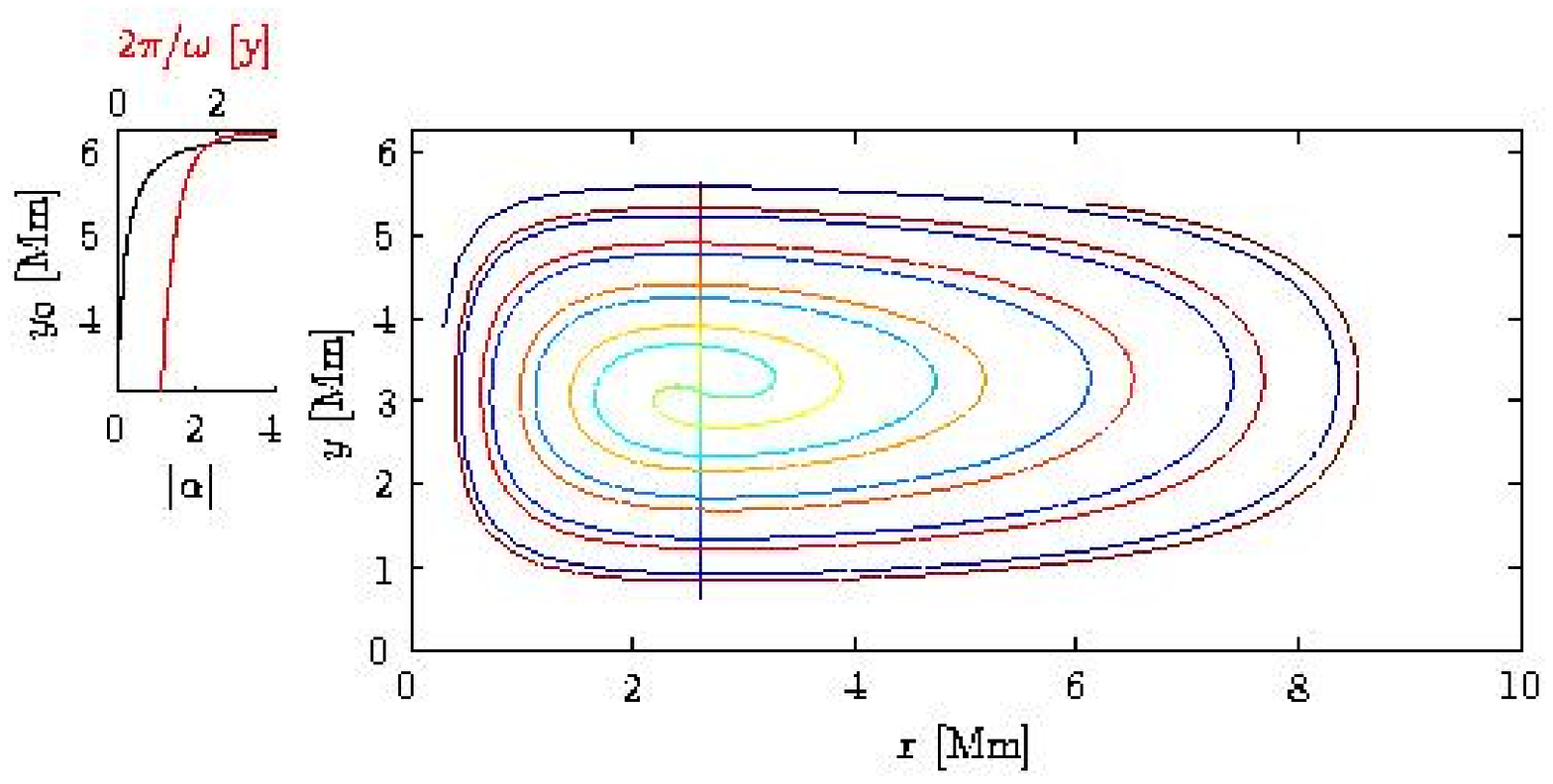}}\quad
\subfigure[]{\includegraphics[width=7cm,clip=]{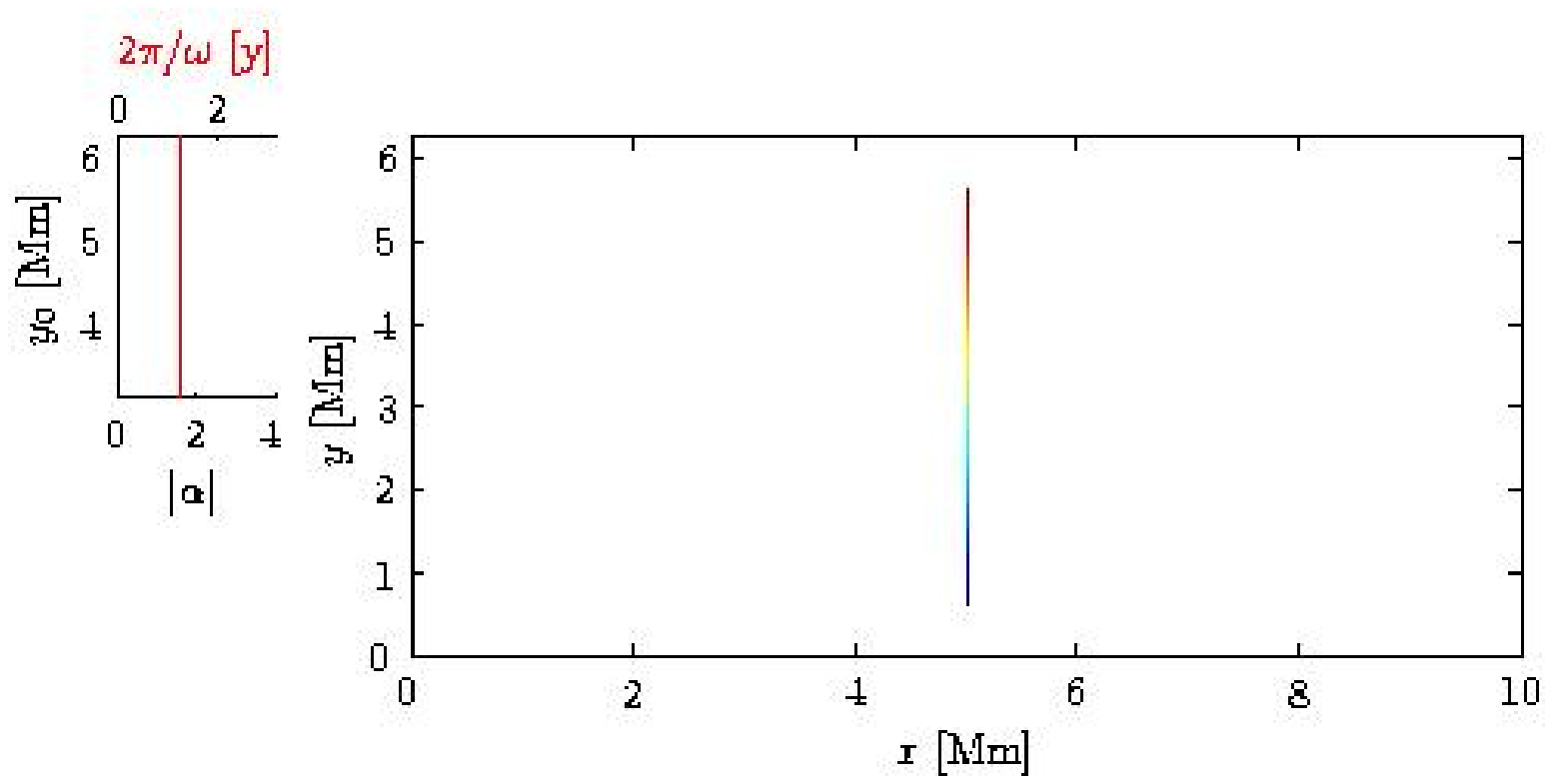}}}
\caption{Evolution of a material line of fluid in background flows
S (a) and R (b). Upper-left plots depict rotational period,
$2\pi/\omega$, and stability parameter, $\alpha$, as a function of
initial meridional position.\label{she}}
\end{figure*}%

A simple but very important observation follows from the action--angle
formalism. Dependence of both particle (\ref{act}) and variational (\ref{var}%
) equations on the background flow enters only through the function $\omega
(I)$. Equations (\ref{var}) strongly suggests that trajectory stability and $%
\omega ^{\prime }$ are closely linked. The following heuristic argument
explains the mechanism by which $\omega ^{\prime }$ is expected to control
trajectory stability. If one assumes that $\varepsilon $ is small and the
second derivatives of $\bar{\psi}^{(1)}$ are zero-mean random variables,
then when $\omega ^{\prime }=0$ these terms should lead to slow (power-law)
growth of $\delta \vartheta $ and $\delta I.$ If $\left| \omega ^{\prime
}\right| $ is large, this term will cause a rapid growth of $\left| \delta
\vartheta \right| $ for any nonzero $\left| \delta I\right| $. The
perturbation terms will then lead to a mixing of $\left| \delta \vartheta
\right| $ and $\left| \delta I\right| $. The term $\omega ^{\prime }$ will
lead, in turn, to further growth of $\left| \delta \vartheta \right| .$ As
this process repeats itself, both $\left| \delta I\right| $ and $\left|
\delta \vartheta \right| $ are expected to grow rapidly. The role played by $%
\omega ^{\prime }$ in this process is to amplify small perturbations caused
by the second term on the r.h.s. of Eqs. (\ref{var}). Thus when $\varepsilon
$ is small, trajectory instability is expected to be significantly enhanced
when $\left| \omega ^{\prime }\right| $ is large. When $\varepsilon $ is
sufficiently large that the two terms on the r.h.s. of Eqs. (\ref{var}) have
comparable magnitude, the role played by $\left| \omega ^{\prime }\right| $
in amplifying perturbations is expected to be much less important. Increased
trajectory instability should result in larger numerical estimates of
Lyapunov exponents. A dynamical-systems-based argument on the role of $%
\omega ^{\prime }$ in controlling trajectory stability is given below; that
argument is consistent with the above heuristic argument.

The lower panels of Fig. \ref{alp} show the absolute value of the stability
parameter
\citep[][]{Zaslavsky-98,Beron-Brown-03,Beron-Brown-03b}%
\begin{equation}
\alpha (I):=\frac{I}{\omega }\frac{\mathrm{d}\omega }{\mathrm{d}I}
\label{def}
\end{equation}%
as a function of trajectory initial condition; recall that these initial
conditions correspond to variable $y_{0}$ with $x_{0}$ fixed at the gyre
center. Comparison of the middle and lower panels of Fig. \ref{alp}a
suggests that when the perturbation to the background steady flow is weak,
trajectory instability increases, on average, with increasing $\left| \alpha
\right| $. Figure \ref{alp}b shows that for a strong perturbation this trend
is less strong, although the most stable trajectories are clearly those in
the region of the flow where $\left| \alpha \right| $ is small. The
background flow R used to produce Fig. \ref{alp}c was chosen because it has
the property $\alpha =0$ $\forall I$. Because the same perturbation flows
were used to produce Fig. \ref{alp}a and \ref{alp}c the difference between
these figures is entirely due to the difference in the background flows. The
remarkable stability of trajectories in Fig. \ref{alp}c is due to the
property $\alpha =0$ $\forall I$ in flow R. The same comment applies to the
difference between Fig. \ref{lag}a and \ref{lag}c, which were produced using
the same flows that were used to produce Fig. \ref{alp}a and \ref{alp}c. All
of the aforementioned observations relating to Figs. \ref{alp} and \ref{lag}
are consistent with the heuristic argument given in the preceding paragraph
describing how $\left| \omega ^{\prime }\right| $ is expected to control
trajectory stability.

The physical interpretation of the stability parameter $\alpha $ is
illustrated in Fig. \ref{she}. Figures \ref{she}a,b show the evolution of a
material line of fluid in background flows S and R, respectively. The
material line is shown at $t=0$ and at $t=12$ \textrm{y.} Also shown is a
plot of both $2\pi /\omega $ and $\left| \alpha \right| $ as a function of $%
y_{0}$ (for $y_{0}>y_{\mathrm{C}},$ the meridional coordinate of the center
gyre) in each environment. As a consequence of the uniqueness of solutions
to Eq. (\ref{sys}) and continuity of the velocity field, the material line
of fluid cannot break or intersect itself but it can increase in complexity
with time. Because the motion in Fig. \ref{she} is integrable (i.e., each
point of the material line is constrained to lie on a surface of constant $I$
$\forall t$) and because attention is restricted to background flows for
which $\psi ^{(0)}$ has compact and closed level sets, i.e., gyre flows, the
length of the material line can grow with time, at most, following a power
law. Background flow R has a special property. In that background flow the
material line just rotates clockwise at a constant rate $\omega =\omega _{%
\mathrm{R}}$ $(=2\pi $ $\mathrm{y}^{-1})$, independent of $I$, so $\alpha =0$
$\forall I.$ In contrast, $\omega $ varies with $I$ in background flow S.
The monotonic decay of $\omega $ as a function of $I$ in background flow S
induces a\textit{\ }shear in phase space which causes the outermost points
of the material line to rotate more slowly than the innermost ones and,
hence, causes the material line to spiral. In background flow R there is no
shear. In polar coordinates radial shear can be defined as
\begin{equation}
r\partial _{r}\left( r^{-1}u_{\theta }\right) ,  \label{rad}
\end{equation}%
where $r$ is the radial coordinate and $u_{\theta }$ is the $\theta $%
-component of the velocity field. More correctly, this quantity is twice the
$r\theta $-component of the strain-rate tensor for rotational motion
\citep[cf., e.g.,][]{Batchelor-64}%
. The connection with motion in phase space can be accomplished by
identifying $I$ with $r$ and $\omega I$ with $u_{\theta }$. The replacements
$r\mapsto I$ and $u_{\theta }\mapsto \omega I$ in (\ref{rad}) thus give the
analogous expression $I\omega ^{\prime }$ for the shear in phase space.
Notice that this expression is (apart from the $\omega ^{-1}$-factor) the
stability parameter $\alpha $. We have chosen to include the $\omega ^{-1}$%
-factor in the definition of $\alpha $ because of precedent
\citep[][]{Zaslavsky-98,Beron-Brown-03,Beron-Brown-03b}
and because it is convenient to make $\alpha $ dimensionless.

To see the importance of the shear in the background flow, compare Figs. \ref%
{she}a,b with Figs. \ref{lag}a,c, which show the evolution of the same
initial material line segments in the total (background plus perturbation)
flows. Notice the highly complicated structure of the segment in the
perturbed flow S (Fig. \ref{lag}a) as compared to that in the unperturbed
one (Fig. \ref{she}a). (Note that the number of particles used to produce
Fig. \ref{lag}a is far too small to resolve what should be an unbroken
smooth curve which does not intersect itself.) In contrast, observe that in
environment R the perturbation has only a very minor effect on the evolution
of the material line (Figs. \ref{lag}c and \ref{she}b).

Additional insight into why $\alpha $ should be expected to control
trajectory stability comes from the following argument. The perturbation
streamfunction $\varepsilon \psi ^{(1)}$ has the effect of introducing
perturbations to the action $I$ of a given particle by the amount $\delta I$%
. If $\delta I$ is assumed to be small and of the same order as the
perturbation streamfunction [$\delta I=O(\varepsilon ),$ say], then $\omega $
experiences the change
\begin{equation*}
\omega \mapsto \left( 1+\alpha \delta I/I\right) \omega
\end{equation*}%
$+$ $O(\varepsilon ^{2}).$ The perturbation to $\omega $ depends on both the
perturbation $\delta I$ and the background flow via $\alpha $. Under the
change $I\mapsto I+\delta I,$ a sufficient condition for $\omega $ to remain
invariant at $O(\varepsilon )$ is $\alpha =0.$ This provides an explanation
for the remarkable stability of the particle trajectories in flow R. To $%
O(\varepsilon )$ a nonvanishing shear $(\alpha \neq 0)$ appears as a
necessary condition to sustain the successive stretching and folding of the
material line of fluid after it gets distorted by the perturbation. (Of
course, chaotic motion is still possible when $\alpha =0$ provided that $%
\varepsilon $ is sufficiently large.) It is thus expected that where $%
|\alpha |$ is small (resp., large) there will be less (resp., more)
sensitivity to initial conditions and, hence, the motion be more regular
(resp., chaotic). Support for this conjecture is given in the numerical
simulations presented in this paper.

Finally, the role of $\omega ^{\prime }$ in dynamical systems theory
deserves further comment. A nondegeneracy condition, $\omega ^{\prime }\neq
0,$ must be satisfied in order for the KAM theorem to apply and, hence, to
guarantee that some trajectories are nonchaotic provided the strength of the
time-dependent perturbation is sufficiently weak. This theorem does not
imply, however, that trajectories are unstable when $\omega ^{\prime }=0;$
the KAM theorem does not address this limit. The mechanism that leads to
chaos is the excitation of resonances at discrete frequencies. For a
sufficiently strong perturbation, neighboring resonances overlap and chaotic
motion results
\citep[cf., e.g.,][]{Tabor-89}%
. The width in frequency of each resonance is proportional to $|\omega
^{\prime }|^{1/2},$ so one expects, on average, motion to become
increasingly chaotic as $\left| \omega ^{\prime }\right| $ increases. This
expected trend is consistent with the arguments and numerical simulations
that we have presented above. The trend toward increasingly chaotic motion
with increasing $\left| \omega ^{\prime }\right| $ does not, of course, rule
out some nonchaotic motion for fixed but large $\left| \omega ^{\prime
}\right| .$ Note also that the trends that we have described apply on
average; details depend on details of the flow, both background and
perturbation.

\section{Concluding remarks\label{con}}

In this paper we considered particle motion in unsteady incompressible
two-dimensional flows consisting of a steady background gyre on which a
highly structured unsteady wave-like perturbation is superimposed. The
numerical simulations presented strongly suggest that: (i) phase space is
mixed, characteristic of near-integrable one-and-a-half-degree-of-freedom
Hamiltonian systems; and (ii) particle trajectory stability strongly depends
on the structure of the background (steady) component of the flow.

The mixed phase space structure, in which ``islands'' of stability emerge
from an otherwise chaotic ``sea,'' was explained as a consequence of the
applicability of KAM theory. The mixed phase space provides an explanation
for the occurrence of patches of poorly stirred fluid in a mostly vigorously
stirred flow. Trajectory instability was shown to increase with increasing
magnitude of $\alpha :=I\omega ^{\prime }/\omega ,$ where $2\pi /\omega (I)$
is the period of revolution of a particle in the background gyre flow and $I$
is the particle's action variable in the background flow.

These results provide important insight into the physics underlying
Lagrangian ocean dynamics. In addition to this insight, the results
described are potentially important in a variety of practical problems. The
occurrence of Lagrangian ``islands of stability'' has important implications
for the transport and dispersal of tracers ranging from nutrients to toxic
pollulants. Knowledge that such ``islands'' are smaller and less abundant,
on average, in regions of flows where $\left| \omega ^{\prime }\right| $ is
large might be exploited when deciding where to place a sewage outfall, for
example.

\begin{acknowledgement}
The comments of an anonymous reviewer have led to improvements in the
manuscript. This work has been supported by Code 321OA of the US Office of
Naval Research.
\end{acknowledgement}

\end{document}